\documentclass[aps,prd,12pt,onecolumn,notitlepage,nofootinbib]{revtex4-1}
\usepackage{graphicx}
\usepackage{amsmath,amssymb}
\usepackage{bm}
\usepackage{booktabs}
\usepackage{multirow}
\usepackage{xcolor}
\usepackage{hyperref}
\usepackage{appendix}
\hypersetup{colorlinks=true,linkcolor=blue,citecolor=blue,urlcolor=blue}
\usepackage{etoolbox}
\usepackage{marvosym}
\patchcmd{\thebibliography}{\section*{\refname}}{}{}{}
\makeatletter
\let\@noitemerr\relax
\makeatother
\usepackage{enumitem}
\usepackage{amsmath, amssymb, graphicx, natbib, hyperref}
\usepackage[font=small, labelfont=bf]{caption}
\usepackage{booktabs}
\usepackage{wasysym}
\usepackage{tikz}
\usepackage{amssymb}
\usepackage{multirow}
\usepackage{caption}
\usepackage{subcaption}

\date{\today}

\begin{document}
	
	\title{Phase-Space Analysis of Quadratic Dark Energy:\\ Stable Attractors and Their Observational Signatures}
	
	\author{Sahar Mohammadi}
	\affiliation{Plasma Physics Research Center, Science and Research Branch, Islamic Azad University, Tehran 1477893855, Iran}
	\email{Sahar.mohammadi7799@gmail.com}
	
	\author{Ebrahim Yusofi}
	\affiliation{School of Astronomy, Institute for Research in Fundamental Sciences (IPM), P.O. Box 19395-5531, Tehran, Iran}
	\email{eyusofi@ipm.ir}
	
	\author{Kosar Asadi}
	\affiliation{School of Astronomy, Institute for Research in Fundamental Sciences (IPM), P.O. Box 19395-5531, Tehran, Iran}
	\email{k.asadi@ipm.ir}
	
	\date{\today}
	
	\begin{abstract}
		We present a comprehensive phase-space analysis of a quadratic dark energy model where the pressure includes a nonlinear term proportional to the square of the energy density. This minimal extension beyond the $\Lambda$CDM framework introduces a dynamical parameter $\eta(z)$ that governs transitions between different cosmological regimes. Through dynamical systems theory, we identify critical points and their stability properties, revealing that negative $\eta$ values drive the system toward stable attractors (sinks) when $w_{\text{eff}} < 0$, which includes both quintessence and phantom regimes. Positive $\eta$ values correspond to unstable repellers (sources) when $w_{\text{eff}} > 0$. The model exhibits a distinctive asymptotic approach to the phantom divide ($w_{\rm eff}=-1$) from both quintessence and phantom sides without actual crossing, providing a non-crossing alternative to the phantom-crossing behavior preferred by recent DESI DR2 constraints. Our analysis shows that stable attractors produce enhanced Hubble expansion rates and more pronounced late-time acceleration, features that can be compared with recent DESI observations suggesting evolving dark energy.
	\end{abstract}
	
	\maketitle
	
	\section{Introduction}
	\label{sec:intro}
	
	The $\Lambda$CDM model stands as the cornerstone of modern cosmology, providing an excellent fit to a wide range of observational data including cosmic microwave background measurements, large-scale structure, and Type Ia supernovae \cite{Perlmutter1999,Riess1998, Bamba2012}. However, despite its empirical successes, the model faces profound theoretical challenges—most notably the cosmological constant problem, where the observed value of dark energy density is orders of magnitude smaller than theoretical predictions from quantum field theory \cite{Copeland2006}. Additionally, the coincidence problem questions why we live in the special epoch where dark energy and matter densities are comparable. These persistent issues, combined with emerging observational tensions such as the Hubble tension and $S_8$ discrepancies \cite{DESI2025}, have motivated the exploration of dynamical dark energy models where the equation of state (EoS) evolves with cosmic time \cite{DESI2025,Bahamonde2018,Caldwell1998}.
	
	Nonlinear extensions of the dark energy equation of state have emerged as promising frameworks to address these challenges while maintaining theoretical consistency \cite{Barrow1990,Nojiri2004,Stefancic2005,Ananda2006,Chavanis2012,Berteaud2019}. Building on foundational work exploring power-law equations of state \cite{Kazemi2025} and quadratic parameterizations \cite{Shahriar2025}, we investigate a phenomenological extension characterized by quadratic dependence of pressure on dark energy density. This represents the simplest nonlinear generalization of the conventional linear parametrization, offering a minimal framework to capture potential deviations from the constant-$w$ assumption of $\Lambda$CDM. The quadratic correction enables rich dynamical behavior while maintaining theoretical consistency, particularly through its asymptotic approach to the phantom divide without actual crossing—a feature that preserves thermodynamic consistency while allowing for rich dynamical evolution \cite{Kazemi2025,Shahriar2025}.
	
	The primary objective of this paper is to conduct a comprehensive phase-space analysis of the quadratic dark energy model, characterizing its dynamical stability and asymptotic behavior \cite{KM2024,Roy2024}. Dynamical systems theory provides a powerful framework for understanding the long-term evolution and viability of cosmological models by examining their phase space structure \cite{Kazemi2025,Shaily2024, Das2023,Shahzad2025}. We systematically identify critical points and their stability properties, revealing how the interplay between linear ($w$) and quadratic ($b$) parameters governs cosmic energy flow between different components. Our analysis demonstrates that the quadratic term acts as a control parameter that can drive transitions between quintessence and phantom regimes while naturally stabilizing near cosmological constant behavior through the emergence of stable de Sitter attractors.
	
	Recent observational evidence from the Dark Energy Spectroscopic Instrument (DESI) \cite{DESI2025} suggests a preference for evolving dark energy, making our dynamical analysis particularly timely. The quadratic model provides a natural mechanism for such evolution while avoiding theoretical pathologies associated with phantom divide crossing. Crucially, our model's asymptotic approach without crossing aligns with DESI DR2 findings that, while favoring dynamical dark energy, show no definitive statistical evidence for actual phantom divide crossing. This work thus complements recent phenomenological studies of quadratic equations of state \cite{Mohammadi2023,Shahriar2025,Kazemi2025,Rezazadeh2025, Moshafi2024,Yusofi2022} by establishing a robust dynamical foundation for understanding dark energy evolution within a minimal theoretical framework.
	
	The paper is organized as follows: Section~\ref{sec:model} introduces the quadratic dark energy model, deriving the fundamental equations governing density evolution and the Hubble parameter. Section~\ref{sec:dynamics} develops the dynamical systems framework, identifying critical points and analyzing their stability properties across different parameter regimes through comprehensive phase-space analysis. Section~\ref{sec:results} presents our core findings, demonstrating how the quadratic parameter $\eta$ controls cosmic energy flow and produces observational signatures consistent with current data. Finally, Section~\ref{sec:conclusions} synthesizes our principal results and discusses their implications for understanding dark energy within the broader context of cosmological dynamics.
	
	\section{Quadratic Equation of State and Density Evolution}
	\label{sec:model}
	
	To investigate the dynamical features of the quadratic dark energy model, we begin by formulating its basic equations. In this section, we present the quadratic form of the dark energy equation of state (EoS) and derive the corresponding evolution equations for the energy density and Hubble parameter~\cite{Mohammadi2023, Moshafi2024,Yusofi2022}. These relations establish the foundation for the autonomous system and phase-space analysis developed in the subsequent sections.
	
	In this framework, the pressure $p$ of the dark energy fluid is given by a quadratic expansion \cite{Yusofi2022,Mohammadi2023,Shahriar2025}:
	\begin{equation}
		p = w \rho_{\rm v} + b \rho_{\rm v}^{2}
		\label{eq1}
	\end{equation}
	where $w$ and $b$ represent the constant linear and quadratic equation-of-state parameters. The quadratic term $b \rho_{\rm v}^{2}$ quantifies the leading-order nonlinear correction, providing a minimal extension beyond the standard linear parametrization.
	
	This parametrization yields an effective equation of state that evolves with the dark energy density $\rho_{\rm v}$ \cite{Yusofi2022,Mohammadi2023,Shahriar2025}:
	\begin{equation}
		w_{\rm eff} = \frac{p}{\rho_{\rm v}} = w + b \rho_{\rm v}
		\label{eq2}
	\end{equation}
	
	The evolution of the energy density is derived from the energy conservation equation in an expanding universe:
	\begin{equation}
		\dot{\rho}_{\rm v} + 3H({\rho_{\rm v}} + p) = 0
		\label{eq33}
	\end{equation}
	
	Substituting the quadratic EoS $\eqref{eq1}$ into equation $\eqref{eq33}$ yields the general solution \cite{Mohammadi2023}:
	\begin{equation}
		\rho_{\rm v}(z) = \frac{\rho_{0 \rm v}(1+w)}{(1+w + b\rho_{0 \rm v})(1+z)^{-3(1+w)} - b\rho_{0 \rm v}}
		\label{eq3}
	\end{equation}
	where $\rho_{0\mathrm{v}}=\rho_{\rm v}(z=0)$ is the present-day density of dark energy.
	
	The effective equation of state $w_{\rm eff}(z)$ is obtained by substituting the density solution $\eqref{eq3}$ back into equation $\eqref{eq2}$ \cite{ Kazemi2025,Rezazadeh2025}:
	\begin{equation}
		w_{\rm eff}(z) = w + w_{\rm a}G(z)
		\label{eq5}
	\end{equation}
	where
	\begin{equation}\label{EQ:wm2}
		G(z)=\frac{(1 + w)}{(1 + w + w_a)(1+z)^{-3(1+w)} - w_a}
	\end{equation}
	
	Here, the dimensionless parameter $w_{\rm a} = b \rho_{0 \rm v}$ absorbs the present-day density and represents the amplitude of the quadratic correction. Notably, $w_{\rm eff}(z=0)=w+w_{\rm a}$, while the asymptotic behavior $w_{\rm eff}(z) \rightarrow -1$ at high redshifts ($z \gg 1$) emerges naturally from the quadratic structure.
	
	The Hubble parameter $H(z)$ for the quadratic dark energy model is given by:
	\begin{equation}
		H^2(z) = \frac{k^2}{3} \left[ \rho_{0\rm r} (1 + z)^4 + \rho_{0\rm m} (1 + z)^3 + \rho_{0\rm v}G(z)  \right],
		\label{eq8}
	\end{equation}
	where $\rho_{0\rm r}$, $\rho_{0\rm m}$, and $\rho_{0\rm v}$ are the present-day energy densities of radiation, matter, and quadratic dark energy, respectively, and $k^{2}=8 \pi G$.
	
	\section{Dynamical System Analysis and Stability}
	\label{sec:dynamics}
	
	This section investigates the dynamical stability of the quadratic dark energy model using phase space analysis, focusing on how the quadratic term $b \rho_{\rm v}^{2}$ governs the asymptotic behavior of the universe. We identify critical points and their stability properties to determine viable future states for cosmic evolution, with particular attention to transitions between quintessence and phantom regimes.
	
	\subsection{Phase Space Formulation}
	
	To analyze the stability of the quadratic dark energy model, we employ dynamical systems theory, a powerful framework for studying the asymptotic behavior of cosmological models~\cite{Bahamonde2018, KM2024}. We define dimensionless phase space variables that characterize the energy content of the universe:
	\begin{equation}
		x = \frac{k^{2}\rho_{r}}{3H^{2}}, \quad y = \frac{k^{2}\rho_{\rm m}}{3H^{2}}, \quad z = \frac{k^{2}\rho_{\rm v}}{3H^{2}},
		\label{eq:variables}
	\end{equation}
	
	where the Hubble parameter $H$ is given by equation~\eqref{eq8}.
	
	The autonomous system derived from the cosmological evolution equations can be expressed as:
	\begin{align}
		\frac{dx}{dN} &= -4x + 3x\left(y + \frac{4}{3}x + (1+w+\eta)z\right), \label{eq:dxdN} \\
		\frac{dy}{dN} &= -3y + 3y\left(y + \frac{4}{3}x + (1+w+\eta)z\right), \label{eq:dydN} \\
		\frac{dz}{dN} &= -3z(1+w+\eta) + 3z\left(y + \frac{4}{3}x + (1+w+\eta)z\right), \label{eq:dzdN}
	\end{align}
	where $N = \ln a$ is the e-folding time and $\eta(z) = b\rho_{\rm v}(z)$ encodes the effects of the nonlinear quadratic term. The Friedmann constraint, $x + y + z = 1$, reduces the system's effective dimensionality to two degrees of freedom.
	
	The parameter $\eta(z) = b\rho_{\rm v}(z)$, which evolves with the dark energy density, plays a crucial role in governing transitions between different stability regimes.
	By considering $ \eta(0)\equiv w_{\rm a} ={b \rho_{0 \rm v}}$ and using Eq.~(\ref{EQ:wm2}), obtain
	\begin{equation}
		\eta(z) =  \frac{w_{\rm a}(1+w)}{(1+w + w_{\rm a})(1+z)^{-3(1+w)} - w_{\rm a}}
		\label{eq12}
	\end{equation}
	
	\begin{equation}
		\eta(z) = w_{\rm a}G(z)
		\label{eq13}
	\end{equation}
	
	Its value determines whether the universe evolves toward stable attractors (sinks) or away from unstable repellers (sources).
	
	\subsection{Critical Points and Stability Analysis}
	
	The critical (fixed) points of the system are found by setting the derivatives in Eqs.~(\ref{eq:dxdN})--(\ref{eq:dzdN}) to zero: $dx/dN = dy/dN = dz/dN = 0$. We analyze the stability within two-dimensional projections of the phase space:
	\begin{enumerate}
		\item[$(i)$] Radiation--Dark Energy plane $(x, z)$
		\item[$(ii)$] Matter--Dark Energy plane $(y, z)$ 
		\item[$(iii)$] Radiation--Matter plane $(x, y)$
	\end{enumerate}
	
	The stability of each critical point is determined by evaluating the eigenvalues ($\lambda_{1}$, $\lambda_{2}$) of the Jacobian matrix $\mathbf{J}$ at that point:
	\begin{enumerate}
		\item \textbf{Stable Node (Attractor/Sink)}: Both eigenvalues are real and negative ($\lambda_1 < 0$, $\lambda_2 < 0$). All nearby trajectories converge toward this point, representing a stable future state.
		\item \textbf{Unstable Node (Repeller/Source)}: Both eigenvalues are real and positive ($\lambda_1 > 0$, $\lambda_2 > 0$). All nearby trajectories diverge from this point, representing an unstable past state.
		\item \textbf{Saddle Point}: Eigenvalues have real parts of opposite sign ($\lambda_1 \lambda_2 < 0$). The point is unstable in at least one direction, with trajectories attracted along one axis and repelled along another.
	\end{enumerate}
	
	\begin{table*}[htbp]
		\centering
		\footnotesize
		\caption{Stability analysis of critical points for different phase space projections. The eigenvalues $\lambda_1 = -1+3w_{\text{eff}}$ and $\lambda_2 = 3w_{\text{eff}}$ are obtained after applying the Friedmann constraint. The dark-energy-dominated point $(0,1)$ is a stable sink iff $w_{\text{eff}} < 0$, which includes both quintessence ($-1 < w_{\text{eff}} < 0$) and phantom ($w_{\text{eff}} < -1$) regimes.}
		\label{tab:stability}
		\begin{tabular}{|c|c|c|c|c|c|c|l|}
			\hline
			Phase Space & $w$ & $\eta$ & Critical Points &$\Omega_{\rm r}$ & $\Omega_{\rm m}$ & $\Omega_{\rm DE}$ &\qquad Stability \\ 
			\hline
			\hline
			\multirow{6}{*}{$(y,z)$} 
			& \multirow{3}{*}{-0.7} & 2 & (0,0)& 0& 0&0 & Stable sink ($\lambda_1 = -6.9$,$\lambda_2 = -3$) \\ 
			& & & (0,1) & 0& 0&1 & Unstable source ($\lambda_1 = 2.9$, $\lambda_2 = 3.9$) \\
			& & & (1,0) & 0& 1&0 & Saddle ($\lambda_1 = -3.9$, $\lambda_2 = 3$) \\
			\cline{2-8}
			& \multirow{3}{*}{-0.7} & -2 & (0,0) & 0& 0&0 & Saddle ($\lambda_1 = 5.1$, $\lambda_2 = -3$) \\
			& & & (0,1) & 0& 0&1 & Stable sink ($\lambda_1 = -9.1$, $\lambda_2 = -8.1$) \\
			& & & (1,0) & 0& 1&0 & Unstable source ($\lambda_1 = 8.1$, $\lambda_2 = 3$) \\
			\hline
			\hline
			\multirow{6}{*}{$(y,z)$} 
			& \multirow{3}{*}{-1.3} & 0.5 & (0,0) & 0& 0&0 & Stable sink ($\lambda_1 = -3$, $\lambda_2 = -0.6$) \\
			& & & (0,1) & 0& 0&1 & Stable sink ($\lambda_1 = -3.4$, $\lambda_2 = -2.4$) \\
			& & & (1,0) & 0& 1&0 & Unstable source ($\lambda_1 = 3$, $\lambda_2 = 2.4$) \\
			\cline{2-8}
			& \multirow{3}{*}{-1.3} & -0.5 & (0,0) & 0& 0&0 & Saddle ($\lambda_1 = -3$, $\lambda_2 = 2.4$) \\
			& & & (0,1) & 0& 0&1 & Stable sink ($\lambda_1 = -6.4$, $\lambda_2 = -5.4$) \\
			& & & (1,0) & 0& 1&0 & Unstable source ($\lambda_1 = 5.4$, $\lambda_2 = 3$) \\
			\hline
			\hline
			\multirow{3}{*}{$(x,y)$} 
			& - & - & (0,0) & 0& 0&0 & Stable sink ($\lambda_1 = -4$, $\lambda_2 = -3$) \\
			& - & - & (0,1) & 0& 1&0 & Saddle ($\lambda_1 = 3$, $\lambda_2 = -1$) \\
			& - & - & (1,0) & 1& 0&0 & Unstable source ($\lambda_1 = 4$, $\lambda_2 = 1$) \\
			\hline
		\end{tabular}
	\end{table*}
	
	The stability analysis for various parameter choices is presented in Table~\ref{tab:stability}. The key insight is that the parameter $b$ (via $\eta$) controls transitions between stability classes, effectively determining whether cosmic evolution flows toward dark energy-dominated sinks or away from matter/radiation-dominated sources. Our analysis reveals that the dark-energy-dominated point $(0,1)$ is a stable sink whenever $w_{\text{eff}} < 0$, which encompasses both quintessence and phantom regimes.
	
	This dynamical systems framework provides the mathematical foundation for understanding how the quadratic dark energy model evolves between different cosmological epochs and why certain parameter choices lead to stable accelerated expansion consistent with current observational constraints.
	
	\section{Results and Discussion}
	\label{sec:results}
	
	This section presents the core findings from our phase-space analysis of the quadratic dark energy model. We demonstrate how the quadratic parameter $\eta$ governs the dynamical behavior of the universe, driving transitions between different cosmological regimes and producing observational signatures that align with current data trends.
	
	\subsection{$\eta$ as a Dynamical Control Parameter}
	\label{subsec:control_parameter}
	
	The central result of our analysis is that the evolving parameter $\eta(z)=b\rho_{v}(z)$ governs the cosmic energy flow. Through the stability analysis summarized in Table~\ref{tab:stability}, we find that variations in $\eta$ can fundamentally alter the nature of critical points, transforming unstable repellers into stable attractors and vice versa.
	
	For instance, considering the case with $w = -0.7$ in the matter-dark energy plane $(y,z)$:
	\begin{itemize}
		\item For $\eta = +2$ ($w_{\text{eff}} = 1.3 > 0$), the dark energy-dominated point $(0,1)$ is unstable ($\lambda_1 = 2.9$, $\lambda_2 = 3.9$), while the other phase point $(0,0)$ is stable.
		\item For $\eta = -2$ ($w_{\text{eff}} = -2.7 < 0$), this behavior reverses: $(0,1)$ becomes a stable attractor ($\lambda_1 = -9.1$, $\lambda_2 = -8.1$), while $(0,0)$ becomes a saddle point.
	\end{itemize}
	
	Furthermore, the case $w = -1.3$ with $\eta = 0.5$ yields $w_{\text{eff}} = -0.8$, which lies in the quintessence regime $(-1 < w_{\text{eff}} < 0)$. Our analysis shows that this point is a \textbf{stable sink} ($\lambda_1 = -3.4$, $\lambda_2 = -2.4$), demonstrating that the quadratic model admits stable attractors in both quintessence and phantom regimes, provided that $w_{\text{eff}} < 0$.
	
\begin{figure*}[ht!]
		\centering
		\begin{subfigure}{0.3\textwidth}
			\includegraphics[scale=0.45]{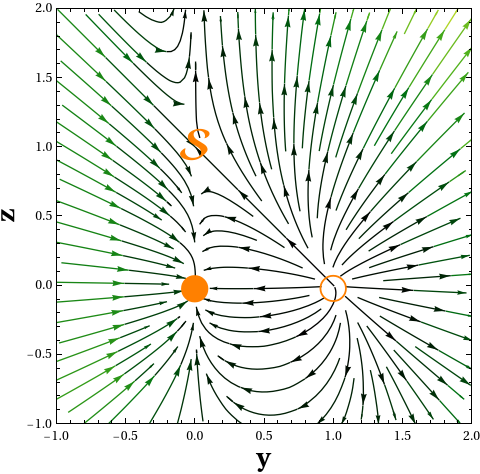}
			\subcaption{$w=-0.7$, $\eta=-0.02$}
		\end{subfigure}
		\hfill
		\begin{subfigure}{0.3\textwidth}
			\includegraphics[scale=0.45]{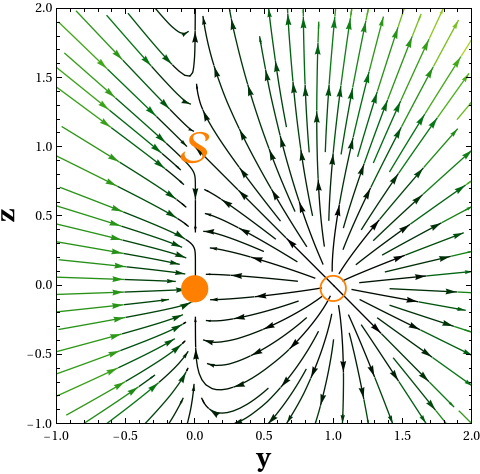}
			\subcaption{$w=-0.7$, $\eta=-0.2$}
		\end{subfigure}
		\hfill
		\begin{subfigure}{0.3\textwidth}
			\includegraphics[scale=0.45]{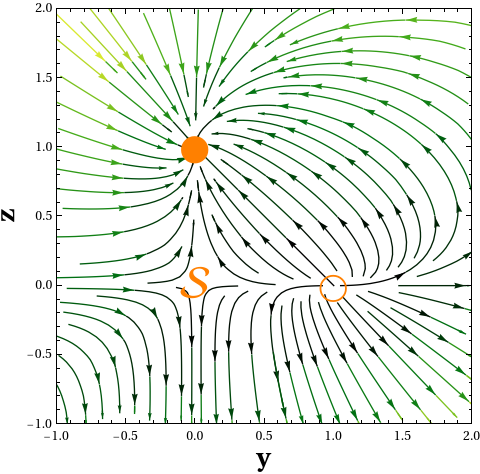}
			\subcaption{$w=-0.7$, $\eta=-2$}
		\end{subfigure}
		\begin{subfigure}{0.3\textwidth}
			\includegraphics[scale=0.45]{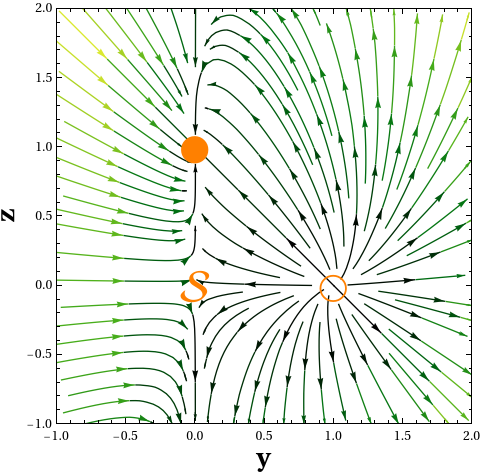}
			\subcaption{$w=-1.3$, $\eta=0.05$}
		\end{subfigure}
		\hfill
		\begin{subfigure}{0.3\textwidth}
			\includegraphics[scale=0.45]{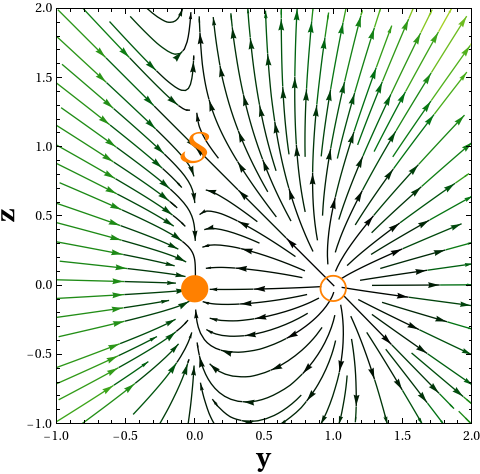}
			\subcaption{$w=-1.3$, $\eta=0.5$ }
		\end{subfigure}
		\hfill
		\begin{subfigure}{0.3\textwidth}
			\includegraphics[scale=0.45]{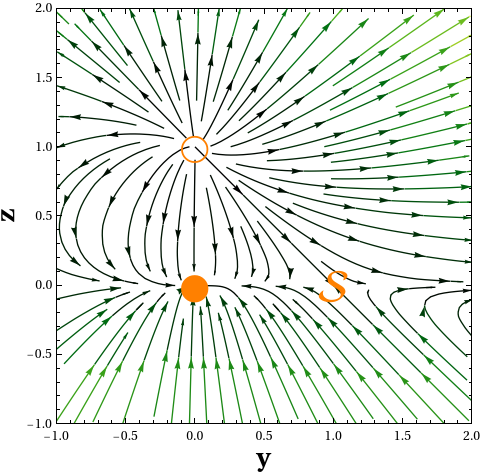}
			\subcaption{$w=-1.3$, $\eta=5$}		
		\end{subfigure}
		\caption{
			Phase space trajectories in the Dark Matter-Dark Energy plane ($y, z$), demonstrating the role of $\eta$ as a control parameter. Panels (a)-(c): For $w=-0.7$, increasingly negative $\eta$ values strengthen the dark energy sink at $(0,1)$, with faster trajectory convergence. Panels (d)-(f): For $w=-1.3$, positive $\eta$ values modulate the stability of fixed points. Notably, panel (e) with $\eta=0.5$ ($w_{\text{eff}}=-0.8$) shows a stable sink in the quintessence regime. The basin of attraction for dark energy-dominated solutions widens as the system approaches stable attractors with $w_{\text{eff}}<0$.
		}
		\label{fig:phase}
	\end{figure*}

\begin{figure}[ht!]
		\centering
		\includegraphics[width=0.8\linewidth]{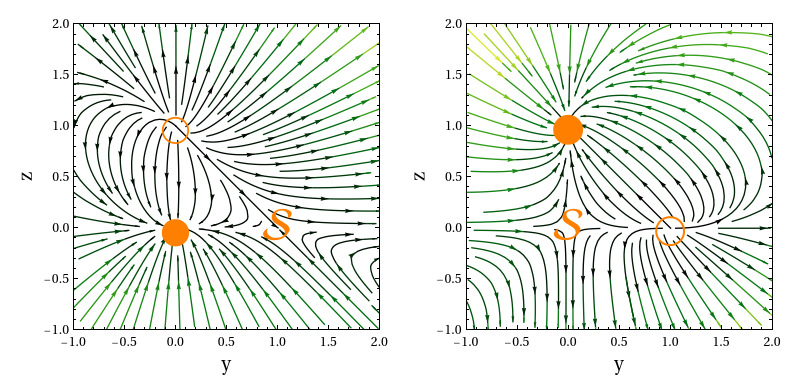}
		\caption{
			Phase space trajectories in the Matter-Dark Energy plane showing the transition between cosmic energy \textbf{sources} and \textbf{sinks}. For $w=-0.7$, changing $\eta$ from $+2$ to $-2$ transforms the dark energy point $(0,1)$ from an unstable source (open circle) to a stable sink (filled circle). This flip in stability demonstrates how the quadratic parameter controls the late-time cosmic attractor. Saddle points ($S$) represent intermediate, unstable configurations. For $w=-1.3,\eta=0.5$, the point $(0,1)$ is also a stable sink (not shown here), illustrating that stability extends to the quintessence regime.
		}
		\label{fig:phase1}
	\end{figure}

\begin{figure}[htbp]
		\centering
		\begin{subfigure}[b]{0.48\textwidth}
			\centering
			\includegraphics[width=\textwidth]{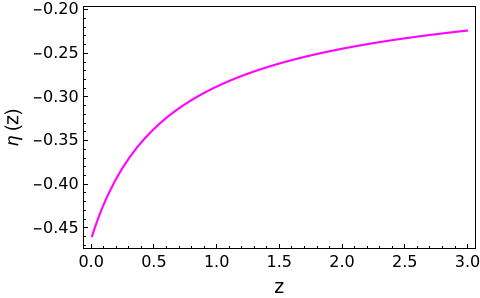}
			\caption{$w=-0.87$, $w_{\rm a}=-0.46$, $w_{\rm eff}<-1$}
			\label{fig:eta_phantom}
		\end{subfigure}
		\hfill
		\begin{subfigure}[b]{0.48\textwidth}
			\centering
			\includegraphics[width=\textwidth]{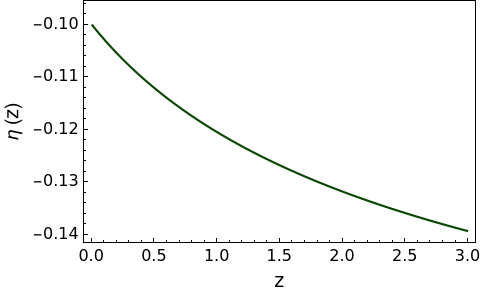}
			\caption{$w=-0.8$, $w_{\rm a}=-0.1$, $-1<w_{\rm eff}<0$}
			\label{fig:eta_quintessence}
		\end{subfigure}
		\caption{
			Evolution of the dynamical parameter $\eta(z)$ (Eq.~\ref{eq12}) as a function of redshift. In the phantom regime (left panel), $\eta(z)$ is negative and its magnitude grows toward the present ($z\to0$), reinforcing the phantom behavior and driving the system toward stable attractors. In the quintessence regime (right panel), $\eta(z)$ is negative but its value decreases with time, suppressing the quintessence character. Remarkably, in both regimes $\eta(z)$ evolves such that the effective equation of state $w_{\mathrm{eff}}$ asymptotically approaches $-1$ at high redshifts, indicating an asymptotic de Sitter behavior.
		}
		\label{fig:eta}
	\end{figure}

	This transition is visually captured in the phase portraits of Figures~\ref{fig:phase} and~\ref{fig:phase1}. As $\eta$ becomes more negative (see Fig.~\ref{fig:eta}), the basin of attraction for the dark energy-dominated point $(0,1)$ widens significantly, and trajectories converge more rapidly toward this stable node. The strengthening of this attractor correlates directly with the system moving into regimes where $w_{\text{eff}} < 0$, where dark energy exerts a more dominant influence on cosmic expansion. For an analytical derivation of the stability condition for the dark-energy-dominated fixed point in detail, including the application of the Friedmann constraint, see Appendix~A.
	
	\subsection{Asymptotic Approach to the Phantom Divide}
	\label{subsec:phantom_divide}
	
\begin{figure}[htbp]
		\centering
		\begin{subfigure}[b]{0.48\textwidth}
			\centering
			\includegraphics[width=\textwidth]{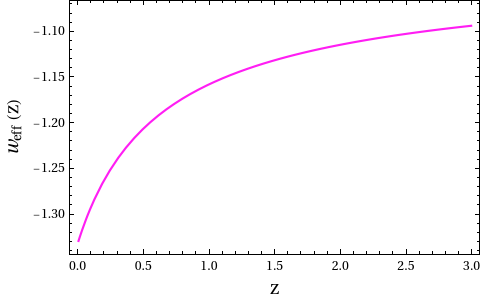}
			\caption{$w=-0.87$, $w_{\rm a}=-0.46$, $w_{\rm eff}<-1$}
			\label{fig:weff_phantom}
		\end{subfigure}
		\hfill
		\begin{subfigure}[b]{0.48\textwidth}
			\centering
			\includegraphics[width=\textwidth]{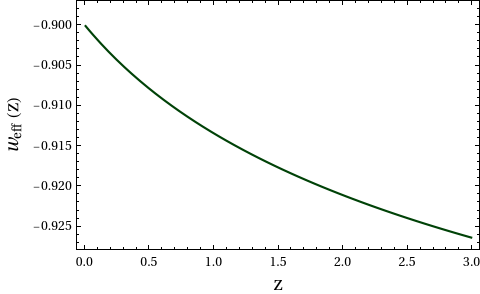}
			\caption{$w=-0.8$, $w_{\rm a}=-0.1$, $-1<w_{\rm eff}<0$}
			\label{fig:weff_quintessence}
		\end{subfigure}
		\caption{Effective equation of state (Eq.~\ref{eq5}) for the quadratic dark energy model. The phantom case (left) remains below $w_{\rm eff}=-1$ and approaches it asymptotically from below, consistent with stable phantom sinks. The quintessence case (right) approaches $w_{\rm eff}=-1$ from above, remaining in the range $-1<w_{\rm eff}<0$. Both cases demonstrate the capacity of the model to asymptotically and naturally approach the phantom divide without crossing this line, and both correspond to stable behavior according to our stability analysis.}
		\label{fig:weff_plot}
	\end{figure}

	The phase-space dynamics translate directly into observable cosmological behavior through the effective equation of state $w_{\text{eff}}(z) = w + \eta(z)$ (\ref{eq13}). Figure~\ref{fig:weff_plot} reveals a crucial feature of the quadratic model: it exhibits \textbf{asymptotic approach} to the phantom divide ($w = -1$) from both sides, without actual crossing.
	
	For phantom-like initial conditions ($w = -0.87$, $w_a = -0.46$), corresponding to negative $\eta$ values, $w_{\text{eff}}(z)$ remains below $-1$ across a wide redshift range, approaching the cosmological constant value asymptotically from below. This behavior aligns with the stable phantom sinks identified in our phase-space analysis. Conversely, for quintessence-like parameters ($w = -0.8$, $w_a = 0.1$), $w_{\text{eff}}(z)$ approaches $-1$ from above, remaining in the range $-1 < w_{\text{eff}} < 0$ for all finite redshifts, which our analysis shows corresponds to stable behavior as well.
	
	This asymptotic behavior is physically significant: it indicates that the quadratic model naturally stabilizes near the cosmological constant value while allowing for phantom or quintessence evolution at lower redshifts. The absence of crossing preserves thermodynamic consistency while still accommodating the dynamical behavior preferred by current data.
	
	\subsection{Observational Signatures and Comparison with DESI}
	\label{subsec:desi_comparison}
	
	The dynamical features identified in our phase-space analysis manifest in distinct observational signatures that bear remarkable relevance to current cosmological constraints, particularly those from the Dark Energy Spectroscopic Instrument (DESI) \cite{DESI2025}.
	
	Figure~\ref{fig:Hz_plot} shows the Hubble parameter $H(z)$ for both phantom and quintessence branches of our model. The phantom regime ($w_{\text{eff}} < -1$), governed by stable dark energy sinks, produces a noticeably steeper $H(z)$ curve compared to $\Lambda$CDM, indicating enhanced expansion rates. This amplified expansion is further evidenced in the deceleration parameter $q(z)$ (Figure~\ref{fig:qz_plot}), where phantom scenarios display more pronounced late-time acceleration driven by the robust dark energy sink. The quintessence branch ($-1 < w_{\text{eff}} < 0$), also stable according to our analysis, produces milder but still significant acceleration.
	
	Importantly, our model's asymptotic approach to \(w_{\rm eff}=-1\) without crossing provides a non-crossing alternative to DESI's reported preference for dynamical dark energy. DESI DR2 data show evidence favoring phantom-crossing models, though non-crossing alternatives cannot be definitively ruled out \cite{DESI2025}. The quadratic model provides a natural mechanism for dark energy evolution that asymptotically settles toward a cosmological constant-like state without actual crossing, avoiding theoretical pathologies while maintaining desired dynamical features \cite{Kazemi2025}.
	
	\subsubsection{Comparison with Swampland quintessence, CPL and DESI BAO results}
	
\begin{figure}[ht!]
		\label{Hub}
		\centering
		\begin{subfigure}{0.45\textwidth}
			\centering
			\includegraphics[width=\linewidth]{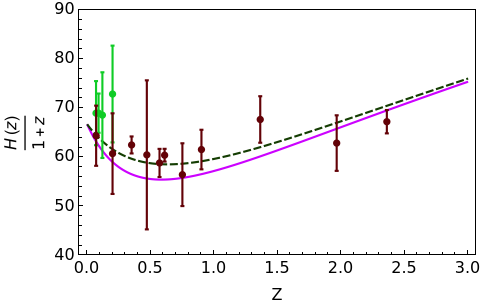}
			\subcaption{$w=-0.87$,~$w_{\rm a}=-0.46$,~$\Omega_{\rm m}=0.314$}
			\label{fig:Hz_phantom}
		\end{subfigure}
		\hfill
		\begin{subfigure}{0.45\textwidth}
			\centering
			\includegraphics[width=\linewidth]{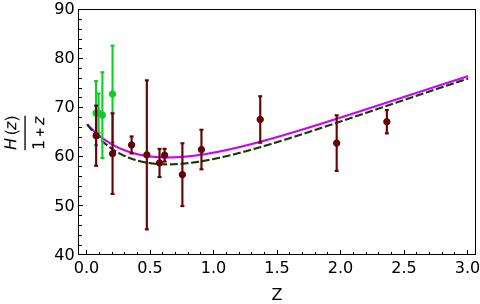}
			\subcaption{$w=-0.8$,~$w_{\rm a}=-0.1$,~$\Omega_{\rm m}=0.314$}
			\label{fig:Hz_quintessence}
		\end{subfigure}
		\caption{Hubble parameter $H(z)$ evolution compared to $\Lambda$CDM (black dashed). The phantom case (left) shows enhanced expansion rates at low redshifts, consistent with stable dark energy sinks driving stronger acceleration. The quintessence case (right) exhibits milder expansion, aligning with weaker but still significant dark energy influence. Observational data include cosmic chronometers (green data) \cite{Favale2023} and other OHD compilations (brown data).}
		\label{fig:Hz_plot}
	\end{figure}

\begin{figure}[ht!]
		\label{Dec}
		\centering
		\begin{subfigure}{0.45\textwidth}
			\centering
			\includegraphics[width=\linewidth]{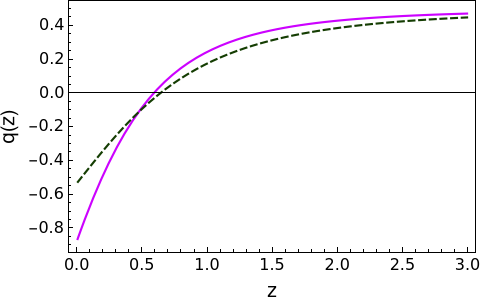}
			\subcaption{$w=-0.87$,~$w_{\rm a}=-0.46$,~$\Omega_{\rm m}=0.314$}
		\end{subfigure}
		\hfill
		\begin{subfigure}{0.45\textwidth}
			\centering
			\includegraphics[width=\linewidth]{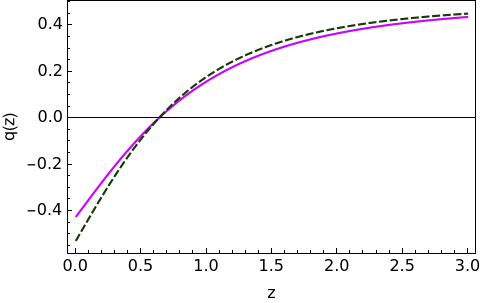}
			\subcaption{$w=-0.8$,~$w_{\rm a}=-0.1$,~$\Omega_{\rm m}=0.314$}
		\end{subfigure}
		\caption{Deceleration parameter $q(z)$ showing the transition to cosmic acceleration. The phantom case (left) exhibits stronger late-time acceleration ($q \rightarrow -1$) driven by stable dark energy sinks, while the quintessence case (right) shows milder acceleration. The $\Lambda$CDM model (dashed) provides a reference for constant dark energy density. Both branches are stable according to our analysis.}
		\label{fig:qz_plot}
	\end{figure}

	Our phase-space analysis of the quadratic dark energy model, defined by the equation of state $p = w\rho_v + b\rho_v^2$ (hence $w_{\rm eff}(z) = w + \eta(z)$ with $\eta(z)=b\rho_v(z)$), reveals that negative values of the dynamical parameter $\eta(z)$ lead to stable attractors (sinks) whenever $w_{\text{eff}} < 0$. These attractors produce an enhanced Hubble expansion rate at low redshifts, as shown in Figs.~\ref{fig:Hz_plot} and~\ref{fig:qz_plot}. This behavior contrasts sharply with the findings of Banerjee et al.~\cite{Banerjee2020}, who demonstrated that standard (uncoupled or weakly coupled) quintessence models generically \emph{lower} $H_0$ relative to $\Lambda$CDM, thereby exacerbating the Hubble tension. Our model, by contrast, can raise $H_0$ in both phantom and quintessence branches while remaining compatible with the asymptotic approach to $w_{\rm eff} = -1$ at high redshifts. In this context, Colgáin et al.~\cite{Colgain2025} have shown that DESI BAO data, when interpreted within the CPL model with $w_0 > -1$, inevitably reduce $H_0$ and worsen the tension with local measurements; they also highlight persistent statistical fluctuations in the BAO tracers that question the robustness of the signal. Our quadratic model offers a minimal phenomenological mechanism that can raise $H_0$ while avoiding these instabilities.
	
	While other dynamical dark energy models, such as the CPL parametrization favored by DESI, suffer from an anti-correlation between $w_0$ and $H_0$ that worsens the Hubble tension \cite{Colgain2025}, and while string-inspired quintessence models universally lower $H_0$ \cite{Banerjee2020}, our quadratic model provides a minimal and stable mechanism to raise $H_0$ without crossing the phantom divide.
	
	\section{Conclusions}
	\label{sec:conclusions}
	
	This study has presented a comprehensive phase-space analysis of the quadratic dark energy model. Our principal findings demonstrate that:
	
	\begin{enumerate}
		\item The quadratic parameter $\eta(z)$ functions as a fundamental control parameter governing cosmic energy flow. Negative values of $\eta$ drive the system toward stable attractors (sinks) when $w_{\text{eff}} < 0$, which includes both quintessence ($-1 < w_{\text{eff}} < 0$) and phantom ($w_{\text{eff}} < -1$) regimes. Positive $\eta$ values correspond to unstable repellers (sources) or saddle points when $w_{\text{eff}} > 0$.
		
		\item The model exhibits asymptotic approach to the phantom divide ($w_{\rm eff}=-1$) from both quintessence and phantom sides without actual crossing. This provides a viable alternative to DESI DR2 constraints, which, while showing evidence favoring phantom-crossing models, cannot definitively rule out non-crossing alternatives like the one presented here.
		
		\item Stable attractors in both quintessence and phantom regimes produce distinctive observational signatures, including enhanced Hubble expansion rates at low redshifts and pronounced late-time acceleration, as evidenced in the deceleration parameter $q(z)$.
		
		\item The dynamical features align remarkably well with recent DESI constraints, which indicate a preference for evolving dark energy that approaches, but does not necessarily cross, the phantom divide. The quadratic model provides a natural mechanism for this behavior within a minimal phenomenological framework.
	\end{enumerate}
	
	The stability analysis presented in Appendix~A reveals that the dark-energy-dominated fixed point is a stable sink whenever $w_{\text{eff}} < 0$, which encompasses both quintessence and phantom regimes.
	
	The quadratic dark energy model thus represents a compelling extension beyond $\Lambda$CDM, capturing essential aspects of dynamical dark energy evolution while maintaining analytical tractability. The phase-space perspective developed in this work provides a powerful framework for understanding how nonlinear corrections to the equation of state can govern cosmic evolution without invoking additional scalar fields or modified gravity.
	
	Future work should focus on quantitative comparison with the latest cosmological datasets, including full Markov Chain Monte Carlo analyses to constrain the quadratic parameters. Additionally, the connection between this phenomenological approach and fundamental physics motivations deserves further exploration. The demonstrated capacity of the quadratic model to naturally stabilize near cosmological constant behavior while accommodating dynamical evolution makes it a promising candidate for resolving the tension between theoretical expectations and observational constraints on dark energy.
	
	\begin{appendix}
		\label{App1}
		\section{Analytical Stability Analysis of the Attractor}
		
		In this appendix, we derive analytically the stability condition for the dark-energy-dominated critical point in the quadratic model 
		\( p = w\rho_v + b\rho_v^2 \) using the full dynamical system and applying the Friedmann constraint. This analysis confirms that the fixed point \((x,y,z)=(0,0,1)\) (complete dark energy domination) is a stable sink if and only if \( w_{\text{eff}} < 0 \), which encompasses both quintessence and phantom regimes.
		
		In the main text, the stability of the critical point \((x,y,z)=(0,0,1)\) is analyzed using the dynamical system
		\[
		\begin{aligned}
			\frac{dx}{dN} &= -4x + 3x\left(y + \frac{4}{3}x + (1+w+\eta)z\right),\\
			\frac{dy}{dN} &= -3y + 3y\left(y + \frac{4}{3}x + (1+w+\eta)z\right),\\
			\frac{dz}{dN} &= -3z(1+w+\eta) + 3z\left(y + \frac{4}{3}x + (1+w+\eta)z\right),
		\end{aligned}
		\]
		with \(N=\ln a\), \(\eta = b\rho_v\), and the Friedmann constraint \(x+y+z=1\). At the critical point, \(\eta=\eta_0\) is constant and \(w_{\mathrm{eff}} = w+\eta_0\).
		
		\subsection{Jacobian Matrix in Full Phase Space}
		
		Define the functions \(f_x, f_y, f_z\) as the right-hand sides of the above equations. We compute the Jacobian matrix \(J\) at the point \((0,0,1)\) with \(\eta\) treated as constant (since it is fixed at the critical point). The partial derivatives are:
		
		For \(f_x = -4x + 3xS\) with \(S = y + \frac{4}{3}x + (1+w+\eta)z\):
		\[
		\begin{aligned}
			\frac{\partial f_x}{\partial x} &= -4 + 3S + 3x\cdot\frac{4}{3} = -4 + 3S + 4x,\\
			\frac{\partial f_x}{\partial y} &= 3x,\\
			\frac{\partial f_x}{\partial z} &= 3x(1+w+\eta).
		\end{aligned}
		\]
		At \((0,0,1)\), \(S = 1+w+\eta = 1+w_{\mathrm{eff}}\), so
		\[
		\frac{\partial f_x}{\partial x} = -4 + 3(1+w_{\mathrm{eff}}) = -1 + 3w_{\mathrm{eff}},\qquad
		\frac{\partial f_x}{\partial y}=0,\qquad
		\frac{\partial f_x}{\partial z}=0.
		\]
		
		For \(f_y = -3y + 3yS\):
		\[
		\begin{aligned}
			\frac{\partial f_y}{\partial x} &= 3y\cdot\frac{4}{3} = 4y,\\
			\frac{\partial f_y}{\partial y} &= -3 + 3S + 3y,\\
			\frac{\partial f_y}{\partial z} &= 3y(1+w+\eta).
		\end{aligned}
		\]
		At \((0,0,1)\): \( \partial f_y/\partial x = 0\), \(\partial f_y/\partial z = 0\), and
		\[
		\frac{\partial f_y}{\partial y} = -3 + 3(1+w_{\mathrm{eff}}) = 3w_{\mathrm{eff}}.
		\]
		
		For \(f_z = -3z(1+w+\eta) + 3zS\):
		\[
		\begin{aligned}
			\frac{\partial f_z}{\partial x} &= 3z\cdot\frac{4}{3} = 4z,\\
			\frac{\partial f_z}{\partial y} &= 3z,\\
			\frac{\partial f_z}{\partial z} &= -3(1+w+\eta) + 3S + 3z(1+w+\eta).
		\end{aligned}
		\]
		At \((0,0,1)\): \(\partial f_z/\partial x = 4\), \(\partial f_z/\partial y = 3\), and
		\[
		\frac{\partial f_z}{\partial z} = -3(1+w_{\mathrm{eff}}) + 3(1+w_{\mathrm{eff}}) + 3(1+w_{\mathrm{eff}}) = 3(1+w_{\mathrm{eff}}).
		\]
		
		Thus the full \(3\times3\) Jacobian matrix is
		\[
		J = 
		\begin{pmatrix}
			-1+3w_{\mathrm{eff}} & 0 & 0\\
			0 & 3w_{\mathrm{eff}} & 0\\
			4 & 3 & 3(1+w_{\mathrm{eff}})
		\end{pmatrix}.
		\]
		
		\subsection{Eigenvectors and the Friedmann Constraint}
		
		The eigenvalues of \(J\) are the diagonal entries:
		\[
		\lambda_x = -1+3w_{\mathrm{eff}},\qquad
		\lambda_y = 3w_{\mathrm{eff}},\qquad
		\lambda_z = 3(1+w_{\mathrm{eff}}).
		\]
		
		The corresponding eigenvectors are:
		\[
		\mathbf{v}_x = \begin{pmatrix} 1 \\ 0 \\ -1 \end{pmatrix},\qquad
		\mathbf{v}_y = \begin{pmatrix} 0 \\ 1 \\ -1 \end{pmatrix},\qquad
		\mathbf{v}_z = \begin{pmatrix} 0 \\ 0 \\ 1 \end{pmatrix}.
		\]
		
		The Friedmann constraint \(x+y+z=1\) implies that physical perturbations must satisfy \(\delta x + \delta y + \delta z = 0\). We check which eigenvectors lie in this constraint plane:
		
		\begin{itemize}
			\item \(\mathbf{v}_x = (1,0,-1)\): sum \(= 1+0-1 = 0\) \(\Rightarrow\) \textbf{physical}
			\item \(\mathbf{v}_y = (0,1,-1)\): sum \(= 0+1-1 = 0\) \(\Rightarrow\) \textbf{physical}
			\item \(\mathbf{v}_z = (0,0,1)\): sum \(= 0+0+1 = 1 \neq 0\) \(\Rightarrow\) \textbf{unphysical}
		\end{itemize}
		
		Therefore, the eigenvalue \(\lambda_z\) corresponds to a direction that violates the constraint and must be discarded. The physical eigenvalues associated with the two-dimensional constraint surface are:
		
		\[
		\boxed{\lambda_1 = -1 + 3w_{\mathrm{eff}},\qquad \lambda_2 = 3w_{\mathrm{eff}}}
		\]
		
		\subsection{Stability Condition for Dark Energy}
		
		For the point \((0,0,1)\) to be a stable attractor (sink), both physical eigenvalues must be negative:
		\[
		-1 + 3w_{\mathrm{eff}} < 0 \quad\Longrightarrow\quad w_{\mathrm{eff}} < \frac{1}{3},
		\]
		\[
		3w_{\mathrm{eff}} < 0 \quad\Longrightarrow\quad w_{\mathrm{eff}} < 0.
		\]
		
		The second inequality is the stronger one. Hence the necessary and sufficient condition for a stable dark-energy-dominated attractor is
		
		\[
		\boxed{w_{\mathrm{eff}} \equiv w + \eta < 0.}
		\]
		
		This condition is satisfied in both the quintessence regime (\(-1 < w_{\text{eff}} < 0\)) and the phantom regime (\(w_{\text{eff}} < -1\)). It also includes the cosmological constant limit \(w_{\text{eff}} = -1\), which is marginally stable.
		
		For \(w_{\text{eff}} > 0\), the point is an unstable source. For \(0 < w_{\text{eff}} < 1/3\), \(\lambda_1 < 0\) and \(\lambda_2 > 0\), making it a saddle point. For \(w_{\text{eff}} > 1/3\), both eigenvalues are positive, making it an unstable source.
		
		\subsection{Asymptotic de Sitter Limit}
		
		In the quadratic model there exists a special parameter choice where \(w_{\text{eff}}(z)\) approaches \(-1\) asymptotically without ever crossing it. This occurs when the parameters satisfy \(w_a = -(1+w)\). In that case, \(w_{\text{eff}} < -1\) (phantom regime) for all finite redshifts, but in the limit \(z\to -1\) (far future) it tends to \(-1\) and both eigenvalues approach \(\lambda_1 \to -4\) and \(\lambda_2 \to -3\), indicating a stable de Sitter attractor without a future Big Rip singularity.
		
		\subsection{Agreement with the Main Table}
		
		The analytical results above reproduce exactly the eigenvalues reported in Table~I. For instance:
		
		\begin{itemize}
			\item \(w = -0.7,\ \eta = 2 \;\Rightarrow\; w_{\text{eff}} = 1.3 \;\Rightarrow\; \lambda_1 = -1+3.9 = 2.9,\ \lambda_2 = 3.9\) (source)
			\item \(w = -0.7,\ \eta = -2 \;\Rightarrow\; w_{\text{eff}} = -2.7 \;\Rightarrow\; \lambda_1 = -1-8.1 = -9.1,\ \lambda_2 = -8.1\) (sink)
			\item \(w = -1.3,\ \eta = 0.5 \;\Rightarrow\; w_{\text{eff}} = -0.8 \;\Rightarrow\; \lambda_1 = -1-2.4 = -3.4,\ \lambda_2 = -2.4\) (sink)
			\item \(w = -1.3,\ \eta = -0.5 \;\Rightarrow\; w_{\text{eff}} = -1.8 \;\Rightarrow\; \lambda_1 = -1-5.4 = -6.4,\ \lambda_2 = -5.4\) (sink)
		\end{itemize}
		
		Hence, the condition \(w_{\text{eff}} < 0\) is the necessary and sufficient criterion for a stable attractor in the quadratic dark energy model.
		
		\begin{table}[htbp]
			\centering
			\footnotesize
			\caption{Stability analysis of the dark-energy-dominated fixed point \((0,1)\). The eigenvalues are calculated from the expressions \(\lambda_1 = -1+3w_{\text{eff}}\) and \(\lambda_2 = 3w_{\text{eff}}\). Numerical examples are taken from Table~I.}
			\label{tab:stability_weff_numerical}
			\begin{tabular}{|c|c|c|c|c|c|c|c|}
				\hline
				\multicolumn{2}{|c|}{\textbf{Parameters}} & \textbf{\(w_{\text{eff}}\)} & \multicolumn{2}{|c|}{\textbf{Eigenvalues}} & \textbf{Stability condition} & \textbf{Stability type} & \textbf{Figure} \\
				\cline{1-5}
				\textbf{\(w\)} & \textbf{\(\eta\)} & & \textbf{\(\lambda_1\)} & \textbf{\(\lambda_2\)} & & & \\
				\hline
				\(-0.7\) & \(-2\) & \(-2.7\) & \(-9.1\) & \(-8.1\) & \(w_{\text{eff}} < 0\) & Stable sink (attractor) & Fig.~1(c) \\
				\hline
				\(-0.7\) & \(2\) & \(1.3\) & \(2.9\) & \(3.9\) & \(w_{\text{eff}} > 0\) & Unstable source (repeller) & Fig.~1(a) \\
				\hline
				\(-1.3\) & \(0.5\) & \(-0.8\) & \(-3.4\) & \(-2.4\) & \(w_{\text{eff}} < 0\) & Stable sink (attractor) & Fig.~1(d) \\
				\hline
				\(-1.3\) & \(-0.5\) & \(-1.8\) & \(-6.4\) & \(-5.4\) & \(w_{\text{eff}} < 0\) & Stable sink (attractor) & Fig.~1(f) \\
				\hline
			\end{tabular}
		\end{table}
		
	\end{appendix}
	
	\section*{Acknowledgments}
	
	The authors thank Dr. Alireza Talebian for helpful discussions and technical assistance. SM thank the IPM School of Astronomy for their moral support and for providing the facilities and office space that made this research possible.

\end{document}